\documentclass[final,5p,times,twocolumn,number,sort&compress]{elsarticle}

\usepackage{graphicx}
\usepackage{subcaption}
\usepackage{amsmath,amssymb,amsfonts,amsthm,mathtools,bm}
\usepackage{mathrsfs}
\usepackage[scr=rsfs,cal=boondox]{mathalfa}
\usepackage{booktabs}
\usepackage{siunitx}
\usepackage{enumitem}
\usepackage[dvipsnames]{xcolor}
\usepackage{hyperref}
\usepackage{dblfloatfix}
\IfFileExists{orcidlink.sty}{\usepackage{orcidlink}}{\newcommand{\orcidlink}[1]{}}

\hypersetup{
    bookmarks=true,
    colorlinks=true,
    linkcolor=Blue,
    citecolor=Blue,
    filecolor=Blue,
    urlcolor=Blue,
    pdfstartview={FitH}
}

\DeclareMathOperator{\arcsinh}{arcsinh}

\newcommand{\bd}{\mathbf{d}}
\newcommand{\ed}{\mathrm{d}}
\newcommand{\ee}{\mathrm{e}}
\newcommand{\m}{\bar{m}}
\newcommand{\bg}{\bm{g}}
\newcommand{\bomega}{\bm{\omega}}
\newcommand{\bR}{\bm{R}}
\newcommand{\bD}{\bm{D}}

\newcommand{\bxi}{\bm{\xi}}

\newcommand{\M}{\mathscr{M}}
\newcommand{\MI}{\mathscr{M}_{\mathrm{I}}}
\newcommand{\MII}{\mathscr{M}_{\mathrm{II}}}
\newcommand{\MIII}{\mathscr{M}_{\mathrm{III}}}
\newcommand{\MIV}{\mathscr{M}_{\mathrm{IV}}}

\newcommand{\tit}{\tilde{t}}
\newcommand{\h}{\mathscr{H}}

\begin{document}

\begin{frontmatter}

\title{Maximal extension of Schwarzschild-like spacetimes in Lorentz gauge theory}

\author[first]{Mohsen Fathi\,\orcidlink{0000-0002-1602-0722}}
\ead{mohsen.fathi@ucentral.cl}
\affiliation[first]{organization={Centro de Investigación en Ciencias del Espacio y Física Teórica (CICEF), Universidad Central de Chile},
            city={La Serena 1710164},
            country={Chile}}

\begin{abstract}
We study the maximal analytic extension of the Schwarzschild-like black hole solution in Lorentz gauge theory. The lapse function is $f(r)=A_0^{-2}-2\m/r$, so the horizon is located at $r_+=2\m A_0^2$ and the non-affinity coefficient of the horizon generator is $\kappa=1/(4\m A_0^4)$. We first analyze the radial null curves in the Schwarzschild-Droste (SD) and ingoing Eddington-Finkelstein (IEF) charts, and then construct the Kruskal-Szekeres (KS) chart adapted to the LGT geometry. The KS extension contains two exterior regions, a black-hole region and a white-hole region. We also present the standard and regular Carter-Penrose (CP) compactifications. The conformal skeleton is Schwarzschild-like, but the physical scale of the horizon, the surface gravity and the constant-radius curves remain controlled by $A_0$. Hence the solution has the same causal topology as Schwarzschild, while it is geometrically inequivalent to it when $A_0\neq1$.
\end{abstract}

\begin{keyword}
Black holes\sep Lorentz gauge theory\sep Causal structure\sep Maximal extension
\end{keyword}

\end{frontmatter}

\section{Introduction}\label{introduction}

Black holes are useful laboratories for understanding the real geometrical content of a gravity theory. In general relativity, the Schwarzschild solution is not only the exterior field of a static compact source. Its maximal extension also shows that the surface $r=2M$ is only a coordinate singularity, while the physical singularity is located at $r=0$. This became clear from the KS construction and from CP compactification, where the complete spacetime contains two exterior regions, a black-hole region and a white-hole region \cite{Kruskal:1960,Szekeres:1960,Penrose:1964,Carter:1966,Hawking:1973uf,Wald:1984rg,Chandrasekhar:1983,ONeill:1983,Poisson:2004,Frolov:1998wf}. For this reason, when a Schwarzschild-like solution appears in a theory beyond the standard metric formulation, it is natural to ask whether the same global extension still exists.

The question is also relevant in the broader program of testing gravity with compact objects. Modified and gauge formulations of gravity can change the local geometry near the horizon, the behavior of null geodesics and the interpretation of observables \cite{Clifton:2011jh,Berti:2015itd,Bambi:2017iyh,Cardoso:2019rvt}. In such cases, studying shadows, lensing or particle motion is not fully complete unless the causal structure of the spacetime is also under control. Recent examples in other non-standard geometries, such as Lorentzian-Euclidean black holes, also show how null geodesics and CP diagrams can be central to understand the physical meaning of a proposed black-hole metric \cite{Capozziello:2025leb}.

Here we study this issue in Lorentz gauge theory (LGT). In this framework, the basic variables are a Lorentz connection and a scalar field in the fundamental representation of the Lorentz group. The metric is obtained as an emergent geometrical object after symmetry breaking \cite{2019CQGra..36f5015W,2023IJGMM..2050040K,Koivisto:2024asr}. Static black-hole solutions in LGT were obtained in Ref.~\cite{Koivisto:2024asr}. They are close to the Schwarzschild form, but they contain an extra constant from the connection sector.

The solution considered in this letter is described by
\begin{equation}
    f(r)=\frac{1}{A_0^2}-\frac{2\m}{r}.
    \label{eq:lapse_intro}
\end{equation}
For $A_0=1$, the Schwarzschild lapse is recovered. For $A_0\neq1$, however, the metric is not just Schwarzschild written with another radial coordinate. The horizon is shifted to
\begin{equation}
    r_+=2\m A_0^2,
    \label{eq:rplus_intro}
\end{equation}
and curvature scalars keep an explicit dependence on $A_0$. Hence $A_0$ changes the physical scale of the horizon and the geometrical interpretation of the areal radius. Some local and observational aspects of this spacetime have already been studied, including shadows and lensing signatures \cite{Ovgun:2025lgt}, and spinning-particle dynamics \cite{Umarov:2025lgt}. The missing part, which we address here, is the maximal extension and its compact causal representation.

%We use the following abbreviations throughout the paper: SD, IEF, KS and CP. After this point we keep only these abbreviations. 
The plan is simple. We first recall the LGT black hole and the horizon properties. We then discuss radial null curves in the SD and IEF charts. After that we introduce the KS coordinates and draw the full extension. Finally, we give the standard and regular CP diagrams. Our result is that the LGT black hole has a Schwarzschild-like causal skeleton, but the geometrical scale attached to that skeleton is controlled by $A_0$. We use units $G=c=1$ and the signature $(-,+,+,+)$.

\section{The LGT black hole and the horizon}\label{sec:solution}

We briefly recall the LGT ingredients which are needed here. The connection 1-form $\bomega^a{}_{b}$ defines the curvature 2-form
\begin{equation}
    \bR^a{}_{b}=\bD\bomega^a{}_{b}
    =\bd\bomega^a{}_{b}+\bomega^a{}_{c}\wedge\bomega^c{}_{b}.
\end{equation}
The purely connection term is topological. The scalar field $\phi^a$ and the covariant derivative $\bD\phi^a$ introduce the dynamical sector. In the notation of Ref.~\cite{Koivisto:2024asr}, the main invariant pieces can be written schematically as
\begin{equation}
I_{(0)}=\int_{\M} \left(g_1\eta_{ab}\eta^{cd}+g_2\epsilon_{ab}{}^{cd}\right)\bR^a{}_{c}\wedge\bR^b{}_{d},
\end{equation}
\begin{multline}
I_{(2)}=g_3\int_{\M}\bD\phi^a\wedge\bD\phi^b\wedge\bR_{ab}\nonumber\\
+g_4\int_{\M}\epsilon_{abcd}\bD\phi^a\wedge\bD\phi^b\wedge\bR^{cd},
\end{multline}
and
\begin{equation}
I_{(4)}=\lambda\int_{\M}\epsilon_{abcd}\bD\phi^a\wedge\bD\phi^b\wedge\bD\phi^c\wedge\bD\phi^d.
\end{equation}
This short review only fixes the origin of the parameter $A_0$. The full field-equation analysis is not repeated, since our aim is the global extension of the resulting exact metric.

The static spherically symmetric line element is
\begin{equation}
    \bg=-f(r)\bd t^2+f(r)^{-1}\bd r^2+r^2\bd\Omega^2,
    \label{eq:metr_1}
\end{equation}
where
\begin{equation}
    \bd\Omega^2=\bd\theta^2+\sin^2\theta\,\bd\phi^2,
    \qquad
    f(r)=\frac{1}{A_0^2}-\frac{2\m}{r}.
    \label{eq:lapse}
\end{equation}
The SD chart is $(t,r,\theta,\phi)$, and $\m=m_S/(8\pi m_P^2)$. The event horizon is the zero of $f(r)$,
\begin{equation}
    r_+=2\m A_0^2.
    \label{eq:rplus}
\end{equation}
Thus $A_0>1$ enlarges the horizon, while $A_0<1$ contracts it.

The solution is not equivalent to Schwarzschild by a harmless rescaling. For example, the Ricci scalar is
\begin{equation}
    R=\frac{2(A_0^2-1)}{A_0^2 r^2},
    \label{eq:ricci_scalar}
\end{equation}
which vanishes only for $|A_0|=1$. Also the Kretschmann scalar reads
\begin{equation}
    \mathcal{K}=R^{\alpha\beta\gamma\delta}R_{\alpha\beta\gamma\delta}
    =\frac{32\m^2}{r^6}+\frac{4}{r^4}\left[1-f(r)\right]^2.
    \label{eq:kretschmann}
\end{equation}
For $|A_0|=1$, this gives $\mathcal{K}_{\rm S}=48\m^2/r^6$. The horizon can also be detected by invariant methods based on scalar polynomial invariants and gradients of invariants \cite{Abdelqader:2014vaa,Page:2015aia,Tavlayan:2020chf}. These facts show that $A_0$ is not a pure coordinate artifact.

\subsection{IEF chart and the scalar $\mathcal{U}$}\label{subsec:U_scalar}

The tortoise coordinate of Eq.~\eqref{eq:metr_1} is
\begin{equation}
    r_* =\int\frac{\ed r}{f(r)}
    =A_0^2\left[r+r_+\ln\left|\frac{r}{r_+}-1\right|\right].
    \label{eq:rstar}
\end{equation}
The outgoing and ingoing radial null curves in the SD chart are
\begin{align}
    \mathcal{L}^{\rm out}_{(u,\theta,\phi)}:&\qquad t=r_*+u,
    \label{eq:Lout}\\
    \mathcal{L}^{\rm in}_{(v,\theta,\phi)}:&\qquad t=-r_*+v,
    \label{eq:Lin}
\end{align}
where $u$ and $v$ are the retarded and advanced null labels.

By promoting $v=t+r_*$ to a coordinate, one obtains
\begin{equation}
    \bg=-f(r)\bd v^2+2\bd v\bd r+r^2\bd\Omega^2.
    \label{eq:metr_v}
\end{equation}
For later use we introduce
\begin{equation}
    \tit=v-r
    =t+(A_0^2-1)r+2\m A_0^4\ln\left|\frac{r}{r_+}-1\right|.
    \label{eq:tit_ingo}
\end{equation}
Then
\begin{equation}
\begin{split}
    \bg=&-f(r)\bd\tit^2+2\left[1-f(r)\right]\bd\tit\bd r  \\
    &+\left[2-f(r)\right]\bd r^2+r^2\bd\Omega^2 .
\end{split}
    \label{eq:metr_tt}
\end{equation}
All metric components are regular at $r=r_+$. Hence the IEF chart extends the SD chart through the future horizon, but it does not yet contain the past horizon.

It is useful to define
\begin{equation}
    \mathcal{U}(\tit,r)=\left(1-\frac{r}{r_+}\right)
    \exp\left[\kappa(r-\tit)\right],
    \label{eq:U}
\end{equation}
where
\begin{equation}
    \kappa=\frac{1}{4\m A_0^4}=\frac12 f'(r_+).
    \label{eq:kappa}
\end{equation}
The surface $\mathcal{U}=0$ is exactly $\h: r=r_+$. The norm of the metric dual of $\bd\mathcal{U}$ is
\begin{equation}
    \nabla^\mu\mathcal{U}\nabla_\mu\mathcal{U}
    =\frac{(A_0^2-1)(r-r_+)^2\left[r+(A_0^2-1)r_+\right]}
    {A_0^6 r_+^4}.
    \label{eq:normU}
\end{equation}
Thus the normal becomes null on $\h$. In the Schwarzschild limit $|A_0|=1$, the same quantity vanishes everywhere in the IEF domain, but for general $A_0$ it vanishes only on $\h$. This is why the $\mathcal{U}$-constant curves are useful to visualize the way the horizon is approached.

The Killing vector $\bxi=\partial_t=\partial_{\tit}$ satisfies $\bg(\bxi,\bxi)=-f(r)$, so it becomes null on $\h$. On this surface one has
\begin{equation}
    \nabla_{\bxi}\bxi\big|_{\h}=\kappa\,\bxi\big|_{\h}.
    \label{eq:nonaffine}
\end{equation}
Since $\kappa\neq0$, the horizon is non-degenerate. This is the usual surface-gravity normalization for a Killing horizon \cite{Bardeen:1973gs,Poisson:2004,Ashtekar:2004cn}. Standard results on non-degenerate Killing horizons and bifurcate extensions then apply \cite{Boyer_geodesic_1969,Racz:1995nh,Wald:1984rg,Chrusciel:2012jk}. In the present case the KS construction below gives the extension explicitly.

\begin{figure*}[t]
    \centering
    \includegraphics[width=0.98\textwidth]{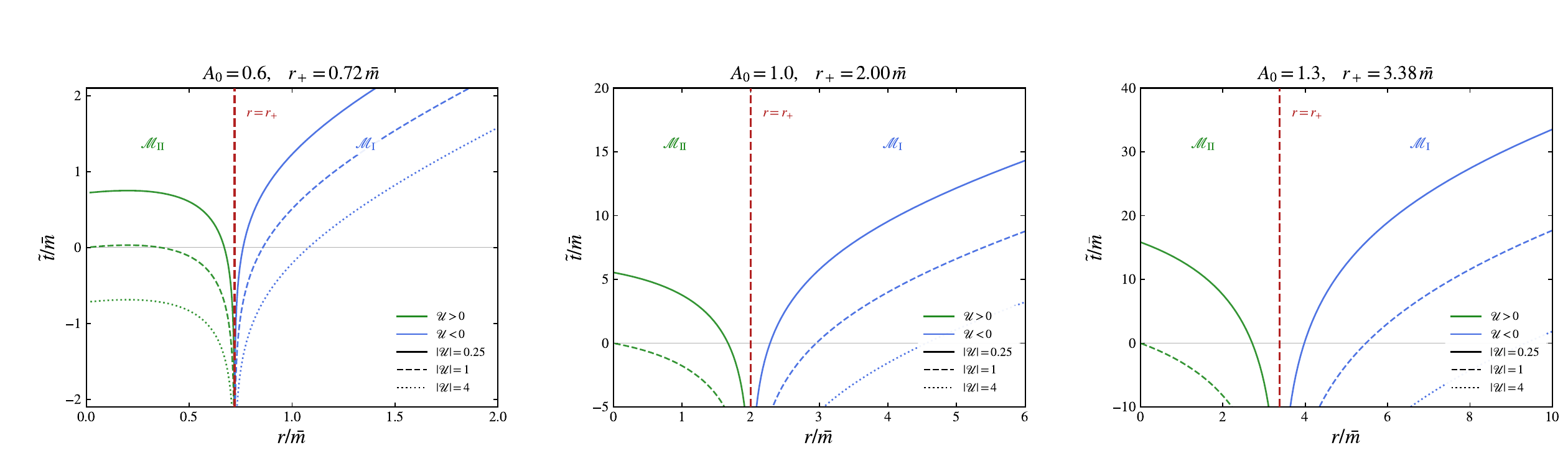}
    \caption{$\mathcal{U}$-constant hypersurfaces in the IEF chart $(\tilde{t},r,\theta,\phi)$ for three representative values of $A_0$. The curves are obtained from $\mathcal{U}=(1-r/r_+)\exp[\kappa(r-\tilde{t})]$. The green curves correspond to $\mathcal{U}>0$ inside the horizon, while the blue curves correspond to $\mathcal{U}<0$ outside the horizon. Different line styles show $|\mathcal{U}|=0.25,1,4$. The red dashed vertical line denotes $r=r_+$.}
    \label{fig:U_constant_IEF}
\end{figure*}

Figure~\ref{fig:U_constant_IEF} shows the $\mathcal{U}$-constant curves. The sign of $\mathcal{U}$ distinguishes the two sides of $\h$: $\mathcal{U}>0$ for $r<r_+$ and $\mathcal{U}<0$ for $r>r_+$.

\subsection{Radial null curves in the SD chart}\label{subsec:SD_null}

The SD chart is useful outside the horizon, but the behavior of Eqs.~\eqref{eq:Lout} and \eqref{eq:Lin} already shows its limitation. Near $r=r_+$ the logarithm in $r_*$ diverges, and the null curves pile up at the horizon in the coordinate picture. In these coordinates the radial null slopes are
\begin{equation}
    \frac{\ed r}{\ed t}=\pm f(r).
    \label{eq:null_slopes_SD}
\end{equation}
Therefore the coordinate speed goes to zero at $r=r_+$, although this is only a coordinate effect.

\begin{figure*}[t]
    \centering
    \includegraphics[width=0.98\textwidth]{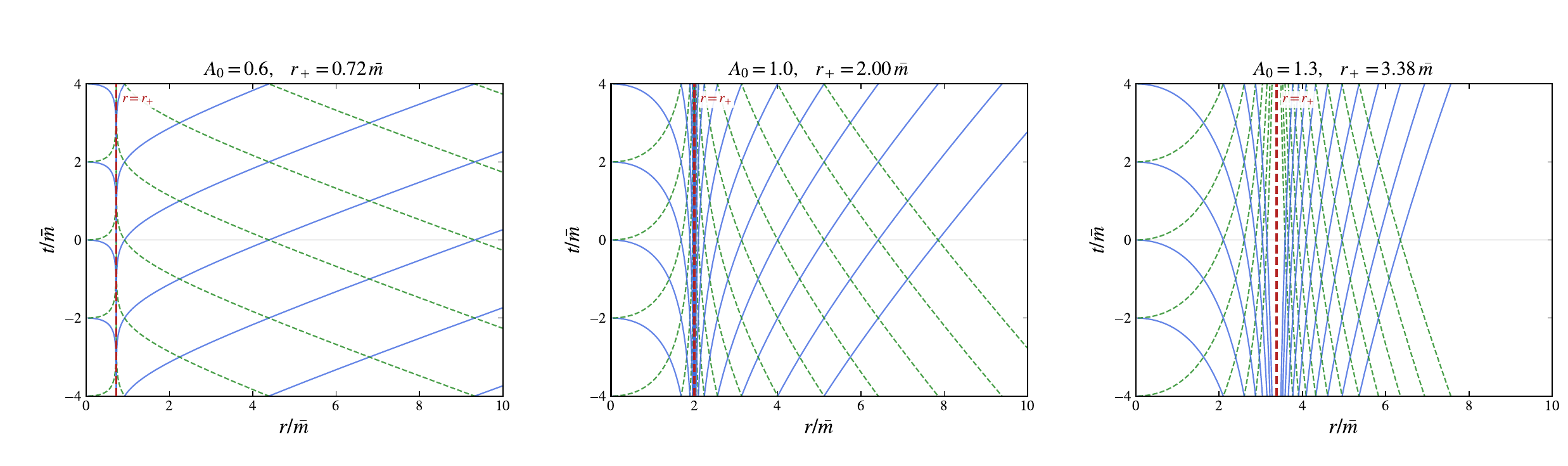}
    \caption{Outgoing and ingoing radial null geodesics in the SD coordinates for three representative values of $A_0$. The solid blue curves represent the outgoing family $t=r_*+u$, while the dashed green curves represent the ingoing family $t=-r_*+v$. The red dashed vertical line marks $r=r_+$. The causal structure is encoded by the two null families, whose intersections determine the local light-cone directions. Near $r=r_+$, the SD chart exhibits the usual coordinate obstruction, which motivates the use of IEF and KS coordinates.}
    \label{fig:SD_null_geodesics}
\end{figure*}

Figure~\ref{fig:SD_null_geodesics} displays the two null families. The solid blue curves are the outgoing family, and the dashed green curves are the ingoing family. For $A_0<1$, the horizon is closer to the origin and the asymptotic value of $f(r)$ is larger. For $A_0>1$, the horizon moves outward and the asymptotic value of $f(r)$ is smaller. This changes the opening of the curves in the coordinate diagram. However, this asymptotic opening is not an invariant statement by itself; it also depends on the normalization of $t$. The invariant information is instead carried by $r_+$, $\kappa$ and the curvature scalars.

\subsection{Radial null curves in the IEF chart}\label{subsec:IEF_null}

For a radial displacement in Eq.~\eqref{eq:metr_tt}, the null condition gives
\begin{equation}
\left(\frac{\ed r}{\ed\tit}\right)^2+
\frac{2[1-f(r)]}{2-f(r)}\frac{\ed r}{\ed\tit}
-\frac{f(r)}{2-f(r)}=0.
\end{equation}
The two branches are
\begin{equation}
    \frac{\ed r}{\ed\tit}=-1,
    \qquad
    \frac{\ed r}{\ed\tit}=\frac{r-r_+}{(2A_0^2-1)r+r_+}.
    \label{eq:IEF_slopes}
\end{equation}
The first family is simply
\begin{equation}
    \mathcal{L}^{\rm in}_{(v,\theta,\phi)}:\qquad \tit=-r+v,
    \label{eq:Lin_IEF}
\end{equation}
and crosses $r=r_+$ smoothly. The second family can be written as
\begin{equation}
    \mathcal{L}^{\rm out}_{(u,\theta,\phi)}:\qquad \tit=2r_*-r+u.
    \label{eq:Lout_IEF}
\end{equation}
It is still singular at the future horizon in the same way expected for outgoing rays. This is exactly what one wants from the IEF chart: it is regular for the future-horizon crossing of ingoing rays.

\begin{figure*}[t]
    \centering
    \includegraphics[width=0.98\textwidth]{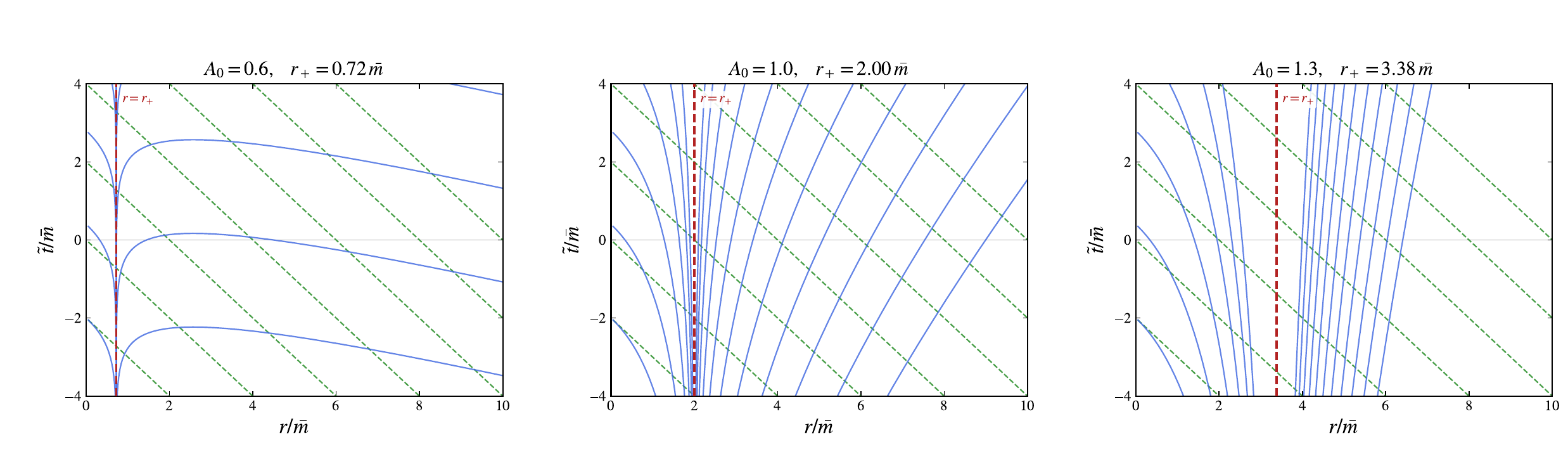}
    \caption{Outgoing and ingoing radial null geodesics in the IEF coordinates for three representative values of $A_0$. The dashed green curves represent $v=\mathrm{constant}$, or $\tilde{t}+r=\mathrm{constant}$, and cross the future horizon smoothly. The solid blue curves represent $u=\mathrm{constant}$, written as $\tilde{t}=2r_*-r+u$. The red dashed vertical line marks $r=r_+$. Compared with the SD chart, the future-horizon crossing of the ingoing null rays is regular.}
    \label{fig:IEF_null_geodesics}
\end{figure*}

This behavior is shown in Fig.~\ref{fig:IEF_null_geodesics}. The dashed green curves pass through $r=r_+$ without obstruction, while the solid blue curves still feel the horizon as a limiting surface in this coordinate patch.

\section{KS extension}\label{sec:KS}

The IEF chart is still not the maximal extension. To see the full geometry, we use the null labels
\begin{equation}
    u=t-r_*,\qquad v=t+r_*.
\end{equation}
The metric becomes
\begin{equation}
    \bg=-f(r)\bd u\bd v+r^2\bd\Omega^2.
    \label{eq:metr_uv}
\end{equation}
The remaining coordinate singularity at $r=r_+$ is removed by defining
\begin{equation}
    U=-\ee^{-\kappa u},\qquad V=\ee^{\kappa v}.
    \label{eq:UV_def}
\end{equation}
These coordinates satisfy
\begin{equation}
    \left(\frac{r}{r_+}-1\right)\ee^{r/r_+}=-UV,
    \label{eq:e(v-u)}
\end{equation}
and the metric takes the regular form
\begin{equation}
    \bg=-\frac{r_+}{\kappa^2 A_0^2 r}\ee^{-r/r_+}\bd U\bd V+r^2\bd\Omega^2.
    \label{eq:metr_UV}
\end{equation}
The coefficient is finite and non-zero at $r=r_+$.

We next introduce
\begin{equation}
    T=\frac{V+U}{2},\qquad X=\frac{V-U}{2}.
    \label{eq:TX_def}
\end{equation}
Then
\begin{equation}
    X^2-T^2=\left(\frac{r}{r_+}-1\right)\exp\left(\frac{r}{r_+}\right),
    \label{eq:KS_hyperbola}
\end{equation}
and
\begin{equation}
    \bg=\frac{r_+}{\kappa^2 A_0^2 r}\ee^{-r/r_+}
    \left(-\bd T^2+\bd X^2\right)+r^2\bd\Omega^2.
    \label{eq:metr_TX}
\end{equation}
In $\MI$, the transformation reduces to
\begin{align}
    T&=\ee^{r/(2r_+)}\sqrt{\frac{r}{r_+}-1}\sinh(\kappa t),
    \label{eq:T_SD_MI}\\
    X&=\ee^{r/(2r_+)}\sqrt{\frac{r}{r_+}-1}\cosh(\kappa t),
    \label{eq:X_SD_MI}
\end{align}
so $T/X=\tanh(\kappa t)$.

\begin{figure*}[t]
    \centering
    \includegraphics[width=0.98\textwidth]{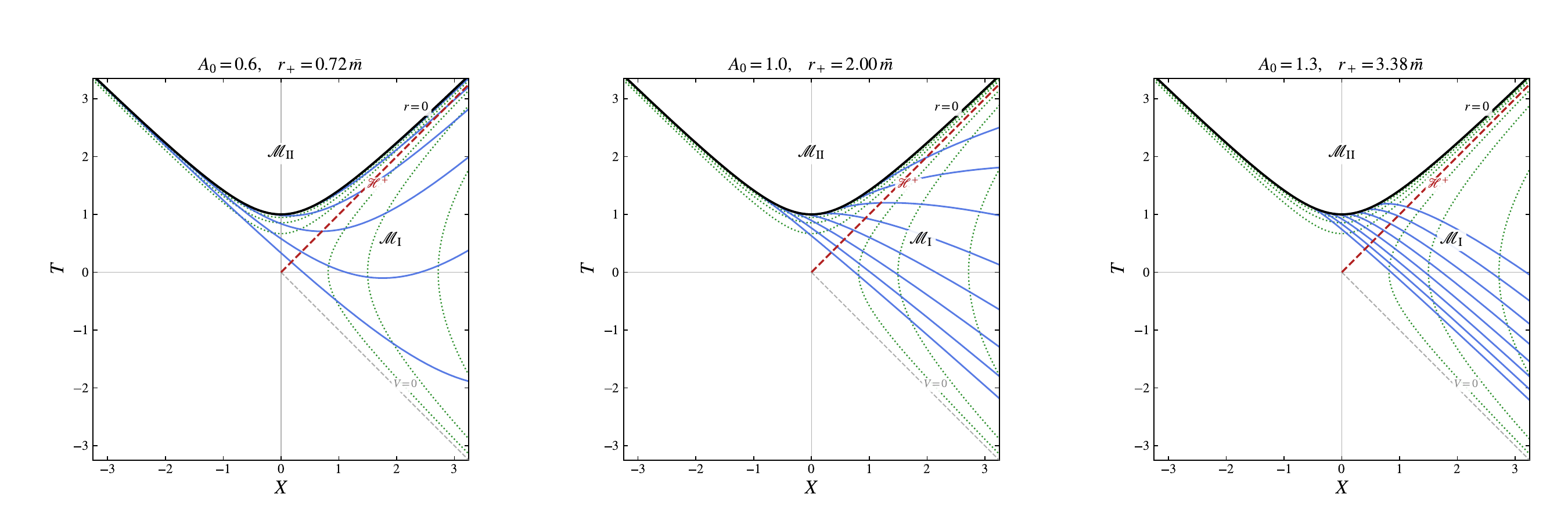}
    \caption{Domain of the IEF coordinates, $\mathscr{M}_{\mathrm{IEF}}$, represented in the KS $(X,T)$ plane for three representative values of $A_0$. The blue solid curves correspond to $\tilde{t}=\mathrm{constant}$ and are spaced by $\Delta\tilde{t}=r_+/2$. The green dotted curves are representative constant-radius curves chosen at $r/r_+=0.20,0.40,0.60,0.80,1.20,1.50,2.00,2.60$. The red dashed line denotes the future horizon $\mathscr{H}^{+}$, the grey dashed line denotes $V=0$, and the black curve is the future singularity $r=0$. The IEF chart covers $\MI$, $\MII$ and the future horizon, but not $\MIII$ or $\MIV$.}
    \label{fig:KS_IEF_titr}
\end{figure*}

The IEF domain inside the KS plane follows from $V=T+X=\ee^{\kappa(\tit+r)}$. Using Eq.~\eqref{eq:e(v-u)}, one obtains
\begin{equation}
    U=T-X=-\left(\frac{r}{r_+}-1\right)
    \exp\left[\frac{r}{r_+}-\kappa(\tit+r)\right].
    \label{eq:U_MIEF}
\end{equation}
Thus the IEF chart covers $T+X>0$ and $T<\sqrt{X^2+1}$. This is only part of the KS plane, as shown in Fig.~\ref{fig:KS_IEF_titr}.

\begin{figure*}[t]
    \centering
    \includegraphics[width=0.98\textwidth]{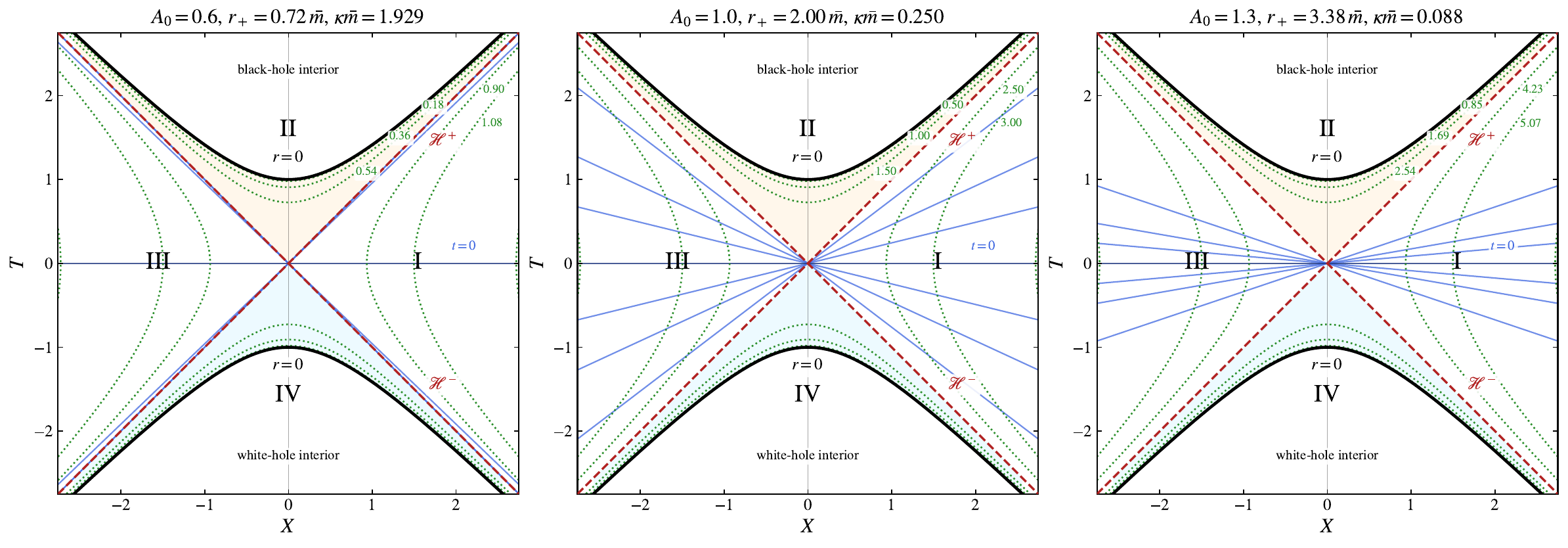}
    \caption{KS diagrams of the LGT Schwarzschild-like black hole for three representative values of $A_0$. The green dotted curves show constant values of $r/\bar m$, while the blue solid rays show constant values of the SD time $t$. The red dashed lines denote the future and past horizons $\mathscr{H}^{+}$ and $\mathscr{H}^{-}$, and the black curves correspond to the curvature singularities at $r=0$. The value $A_0=1$ gives the Schwarzschild limit. For $A_0\neq1$, the causal topology remains the same, but the scale $r_+=2\bar m A_0^2$ and the non-affinity coefficient $\kappa=1/(4\bar m A_0^4)$ change.}
    \label{fig:KS_LGT}
\end{figure*}

The full KS diagram is displayed in Fig.~\ref{fig:KS_LGT}. The radial null directions are $U=\mathrm{constant}$ and $V=\mathrm{constant}$, so they are straight null lines in the $(X,T)$ plane. The horizons are $U=0$ and $V=0$, equivalently $T=X$ and $T=-X$. The singularity is obtained from $r=0$, namely
\begin{equation}
    X^2-T^2=-1.
    \label{eq:singularity_KS}
\end{equation}
The figure makes the extension transparent. The SD chart describes one exterior and the black-hole interior. The IEF chart crosses only the future horizon. The KS chart crosses both null horizons and reveals $\MI$, $\MII$, $\MIII$ and $\MIV$.

\section{Compact CP diagrams}\label{sec:conformalPC}

\subsection{Standard compactification}\label{subsec:standard_CP}

To bring the infinities to finite coordinate distance, we introduce
\begin{equation}
    \hat{V}=\arctan V,
    \qquad
    \hat{U}=\arctan U,
\end{equation}
and
\begin{equation}
    \hat{T}=\hat{V}+\hat{U},
    \qquad
    \hat{X}=\hat{V}-\hat{U}.
    \label{eq:CP_coords}
\end{equation}
Equivalently,
\begin{equation}
    \hat{T}=\arctan(T+X)+\arctan(T-X),
\end{equation}
\begin{equation}
    \hat{X}=\arctan(T+X)-\arctan(T-X).
\end{equation}
The radial null directions remain at $45^\circ$. The horizons are mapped to $\hat{T}=\hat{X}$ and $\hat{T}=-\hat{X}$, while the singularities become the spacelike horizontal boundaries $\hat{T}=\pm\pi/2$.

\begin{figure*}[t]
    \centering
    \includegraphics[width=0.98\textwidth]{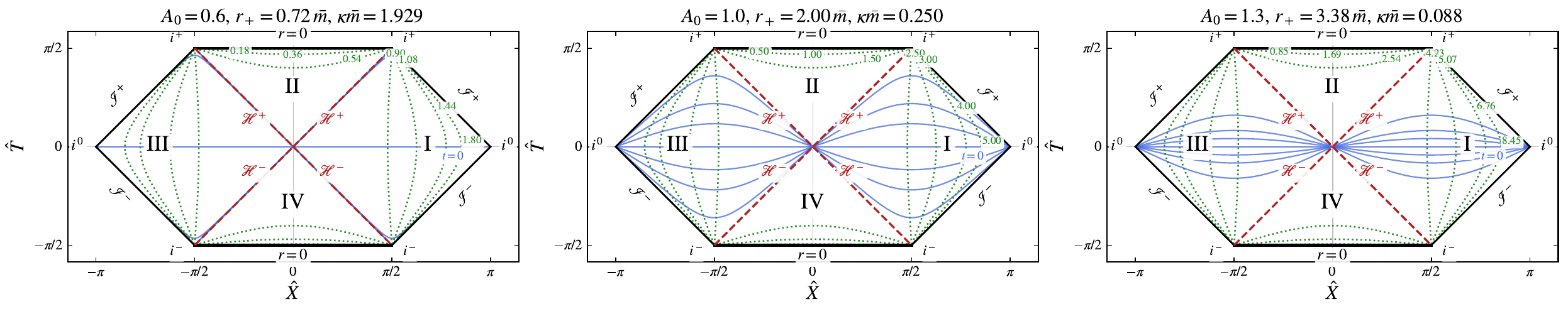}
    \caption{Standard CP compactification of the LGT Schwarzschild-like black hole for three representative values of $A_0$. The green dotted curves show constant values of $r/\bar m$, while the blue curves correspond to constant values of the SD time $t$. The red dashed lines denote $\mathscr{H}^{+}$ and $\mathscr{H}^{-}$, and the black horizontal boundaries correspond to the curvature singularities at $r=0$. The labels $\mathscr{I}^{\pm}$, $i^0$ and $i^{\pm}$ indicate null, spacelike and timelike infinities.}
    \label{fig:CP_standard_LGT}
\end{figure*}

The resulting diagrams are shown in Fig.~\ref{fig:CP_standard_LGT}. They display the same causal topology for all chosen values of $A_0$. What changes is the physical value of $r_+$ assigned to the constant-radius curves and the value of $\kappa$ entering the relation between $t$ and the KS coordinates.

\subsection{Regular compact representation}\label{subsec:regular_CP}

It is also useful to draw the same conformal structure with a smoother compact representation. We use the Penrose-Frolov-Novikov-type map \cite{Frolov:1998wf}
\begin{equation}
    \tilde{V}=\arctan\left[\arcsinh(V)\right],
    \qquad
    \tilde{U}=\arctan\left[\arcsinh(U)\right],
\end{equation}
with
\begin{equation}
    \tilde{T}=\tilde{V}+\tilde{U},
    \qquad
    \tilde{X}=\tilde{V}-\tilde{U}.
    \label{eq:regular_CP_coords}
\end{equation}
This map is monotonic in both null directions, so it preserves the causal character of radial null curves. The condition $X^2-T^2=-1$ is no longer a horizontal line; it is represented by smooth spacelike curves. This does not remove the singularity. It is only a different compact picture of the same spacetime.

\begin{figure*}[t]
    \centering
    \includegraphics[width=0.98\textwidth]{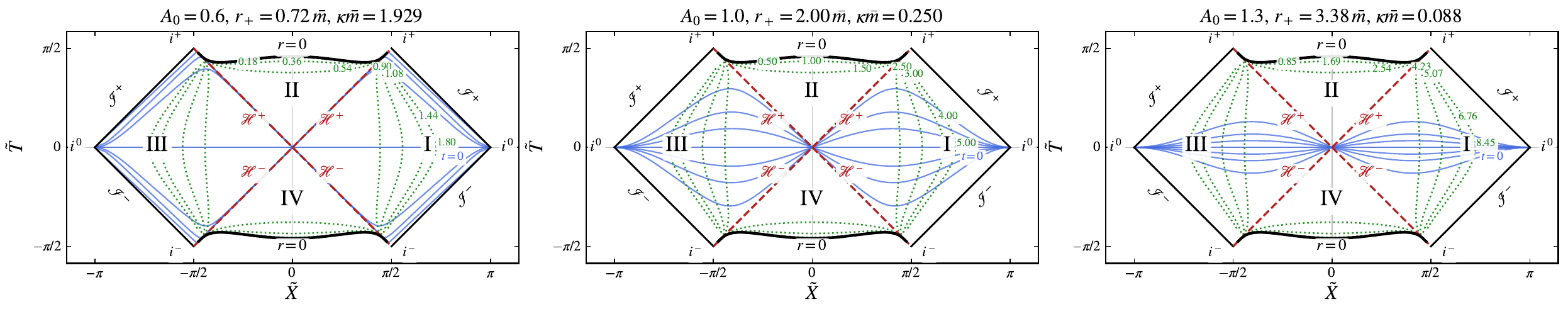}
    \caption{Regular compact CP diagrams of the LGT Schwarzschild-like black hole for three representative values of $A_0$. The compact coordinates are defined through $\tilde{V}=\arctan[\arcsinh(V)]$ and $\tilde{U}=\arctan[\arcsinh(U)]$, with $\tilde{T}=\tilde{V}+\tilde{U}$ and $\tilde{X}=\tilde{V}-\tilde{U}$. The green dotted curves show constant values of $r/\bar m$, while the blue curves correspond to constant values of the SD time $t$. The red dashed lines denote $\mathscr{H}^{+}$ and $\mathscr{H}^{-}$, and the black curved boundaries correspond to the singularities at $r=0$. The singularity is not removed; only its compact representation is changed.}
    \label{fig:CP_regular_LGT}
\end{figure*}

The regular compact diagrams are shown in Fig.~\ref{fig:CP_regular_LGT}. Comparing Figs.~\ref{fig:CP_standard_LGT} and \ref{fig:CP_regular_LGT}, we see that the conformal topology is unchanged. The difference is only the visual form of the compact boundary.

\section{Discussion and conclusions}\label{sec:conclusions}

We have constructed the maximal extension of the Schwarzschild-like black hole solution in LGT. The lapse function $f(r)=A_0^{-2}-2\m/r$ gives the horizon radius $r_+=2\m A_0^2$ and the non-affinity coefficient $\kappa=1/(4\m A_0^4)$. Curvature invariants also depend on $A_0$, so the solution is physically Schwarzschild only for $A_0=1$.

The SD chart shows the usual obstruction at $r=r_+$. The IEF chart removes this obstruction at the future horizon and extends the exterior region into the black-hole interior. However, it does not contain the past horizon, the white-hole region or the second exterior region. These sectors appear naturally after the KS coordinates are introduced.

In the KS chart the metric is regular at the horizon and the relation $X^2-T^2=(r/r_+-1)\exp(r/r_+)$ determines the constant-radius curves. The horizons are $U=0$ and $V=0$, and the singularity is $X^2-T^2=-1$. Therefore the full extension has the same causal arrangement as Schwarzschild: two exterior regions, one black-hole region and one white-hole region.

The CP diagrams confirm this result in compact form. The standard compactification gives horizontal singular boundaries, while the regular compact representation draws the same singularities as smooth spacelike curves. No singularity is removed. The two compactifications only show the same maximally extended spacetime in different visual forms.

The main message is therefore simple. The LGT Schwarzschild-like black hole has a Schwarzschild-type causal topology, but the geometry is not Schwarzschild unless $A_0=1$. The parameter $A_0$ fixes the horizon scale, the surface gravity and the physical meaning of the constant-radius curves. The extension given here provides a clean starting point for future work on thermodynamics, geodesic motion and observational signatures of LGT black holes.

\section*{Acknowledgements}
The author acknowledges financial support from Agencia Nacional de Investigación y Desarrollo (ANID) through the FONDECYT postdoctoral Grant No.~3260029. The author also acknowledges the use of AI tools only for language polishing and improving the clarity of the manuscript.

\bibliographystyle{ieeetr}
\bibliography{1.biblio}

\end{document}